\documentclass[extra, mreferee]{gji}

\usepackage{epsfig}
\usepackage{amssymb, amsmath}
\usepackage{color}
\usepackage{ulem}

\title{Virtual acoustics in inhomogeneous media with single-sided access}

\author[{\small Kees Wapenaar, Joeri Brackenhoff, Jan Thorbecke, Joost van der Neut, Evert Slob, Eric Verschuur}]
{\small Kees Wapenaar, Joeri Brackenhoff, Jan Thorbecke, Joost van der Neut, Evert Slob, Eric Verschuur\\
  Delft University of Technology, P.O. Box 5048, 2600 GA Delft, The Netherlands}

\begin{document}

\label{firstpage}

\def\V{V}
\def\U{\rev{\it \Lambda}}
\def\Vp{\dot V}
\def\S{s}
\def\D{D}
\def\bx{{\bf s}'}
\def\bxb{{\bf x}}
\def\bxr{{\bf r}'}
\def\bxA{{\bf s}}
\def\bxB{{\bf r}}

\def\bx{{\bf x}}
\def\bxr{{\bf x}'}

\def\bxh{\bx_{\rm H}}
\def\bxha{\bx_{{\rm H},A}}
\def\bxhr{\bx_{{\rm H},R}}
\def\setD{\mathbb{V}}
\def\setdD{\mathbb{S}}
\def\setdDR{\setdD_0}
\def\v{v_{\rm n}}
\def\v{V}
\def\pa{p_A}
\def\pb{p_B}
\def\qa{q_A}
\def\qb{q_B} 
\def\half{\begin{matrix}\frac{1}{2}\end{matrix}}

\def\setDA{\mathbb{V}_s}
\def\setdDA{\setdD_s}
\def\xA{x_{3,s}}
\def\chia{\chi_s}

\maketitle

\begin{summary}
{\small A virtual acoustic source inside a medium can be created by emitting a time-reversed point-source response from the enclosing 
boundary into the medium. However, in many practical situations the medium can be accessed from one side only. 
In those cases the time-reversal approach is not exact. Here, we demonstrate the experimental design and use of complex focusing functions to create 
virtual acoustic sources and virtual receivers inside an inhomogeneous medium with single-sided access. 
The retrieved virtual acoustic responses between those sources and receivers mimic the complex propagation and multiple scattering paths of  
waves that would be ignited by physical sources and recorded by physical receivers inside the medium. 
The possibility to predict complex virtual acoustic responses between any two points inside an inhomogeneous medium, without needing a detailed model of the medium, has large
potential for holographic imaging and monitoring of objects with single-sided access, ranging from photoacoustic medical imaging to the
monitoring of induced-earthquake waves all the way from the source to the earth's surface.
}
\end{summary}

\newpage
\section*{Introduction}
In many acoustic applications, ranging from ultrasonics to seismology, 
virtual sources can be created by emitting a  focusing wave field from the boundary into the medium \citep{Porter82JOSA, Oristaglio89IP, Fink92IEEE, Cassereau92IEEE}.
Time-reversal mirroring, developed by Fink and co-workers \citep{Fink92IEEE, Cassereau92IEEE}, is a well-known approach to create a virtual source.
It exploits the fact that the wave equation in a lossless medium is symmetric in time.
In many practical situations,  like in non-destructive testing \citep{Langenberg1986NDT, Saenger2011AMM, Cai2011SMS, Muller2012IJG, Zhu2013SMS},  medical imaging \citep{Tanter2014IEEE, Sapozhnikov2015JASA},
near-field acoustic holography \citep{Maynard85JASA, Wu2004JASA, Wu2016JSV}  
or geophysical holography \citep{Esmersoy88GEO, Lindsey2004AJSS, Gajewski2005GJI}, the medium can be accessed from one side only.
In those cases the time-reversal approach  is not exact, and it breaks down in inhomogeneous  media with strong impedance contrasts.
Recent work by the authors \citep{Wapenaar2013PRL, Slob2014GEO, Wapenaar2016GJI} 
and others \citep{Broggini2014GEO, Meles2015GEO, Ravasi2016GJI, Singh2017GEO2}
 concerns the design of single-sided focusing functions.
  When emitted from the upper boundary into the medium, these focusing functions yield  well-defined foci at predefined positions, which act as omnidirectional virtual sources. 
  This work is inspired by the Marchenko equation of quantum mechanics \citep{Marchenko55DAN, Lamb80Book, Chadan89Book} 
  and its applications in 1D autofocusing \citep{Rose2001PRA, Rose2002IP, Broggini2012EJP}.

We start this paper with a comparison of the time-reversal method and the single-sided focusing approach, at the hand of a number of numerical examples.
Next, we  discuss our approach for retrieving virtual sources and receivers from single-sided reflection data. We apply this methodology to ultrasonic physical model data and seismic reflection data. 
Finally, we discuss potential applications for photoacoustic medical imaging and for monitoring of induced-earthquake waves.

\begin{figure}
\vspace{-5cm}
\centerline{\epsfysize=22 cm \epsfbox{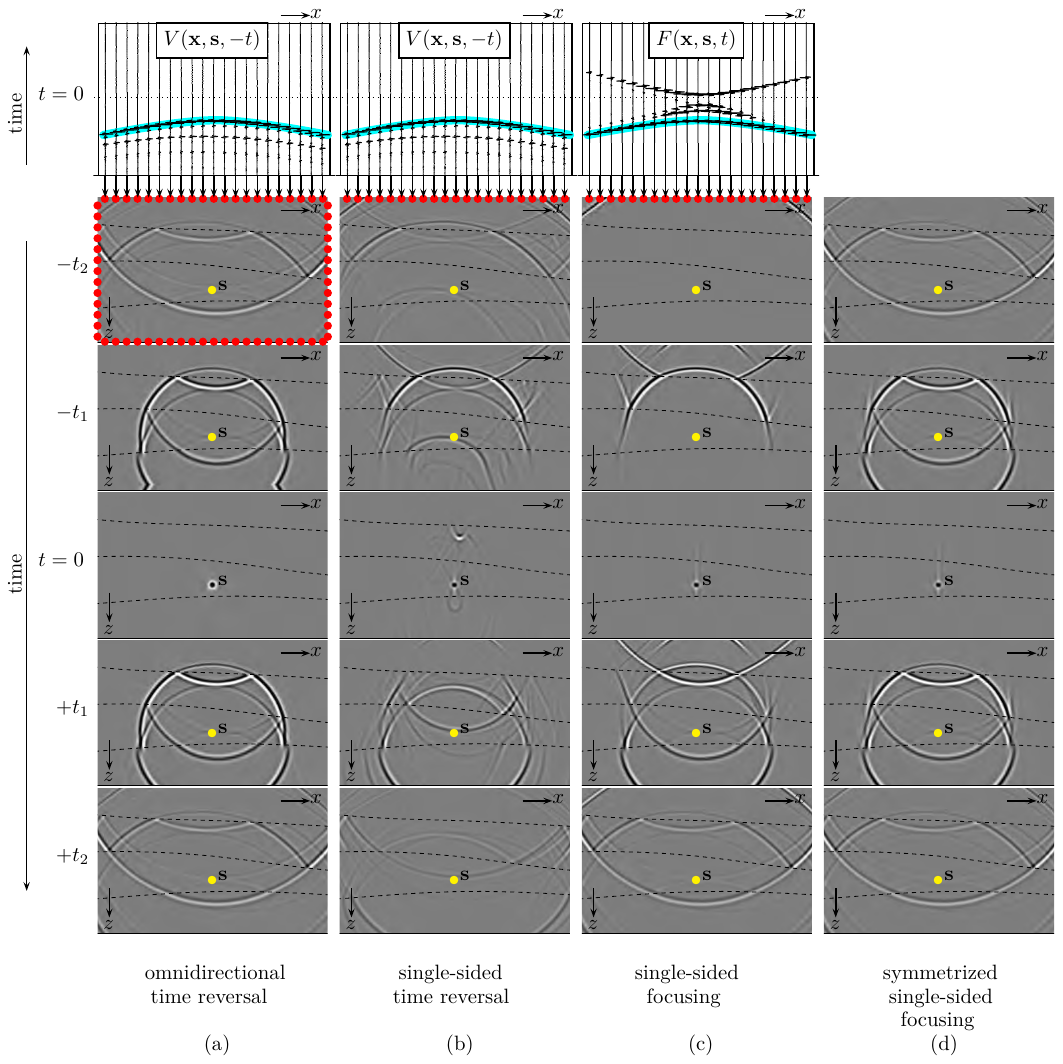}}
\vspace{-3cm}
\caption{\small
Illustration of virtual-source methods. 
(a) 
A time-reversed point source response is emitted from the enclosing boundary into the inhomogeneous medium. 
For negative time, it converges towards the focal point, where it focuses at $t=0$. 
Subsequently, the focal point acts as an omnidirectional radiating virtual source.
(b) 
Emission of the time-reversed response from the upper boundary only. Ghost foci occur at $t=0$. The virtual source radiates mainly downward.
(c) 
Emission of a single-sided focusing function from the upper boundary only. No ghost foci occur at $t=0$. The virtual source radiates mainly downward.
(d) 
Symmetrizing the previous result. No ghost foci occur at $t=0$. The virtual source is omnidirectional.
}\label{Fig1}
\end{figure}

\section*{Time-reversal versus single-sided focusing}
The time-reversal method is illustrated in the first column of Figure \ref{Fig1}, for a lossless layered medium with curved interfaces (denoted by the dashed lines in the grey panels)
 and different propagation velocities and mass densities in the layers between these interfaces.
The top panel shows the time-reversal of the response $\v(\bx,\bxA,t)$ to a point source at $\bxA$ in the third layer of the medium,
as a function of receiver position $\bx=(x,z)$ along the boundary  and time $t$. $\v$ stands for the normal component of the particle velocity.
Only the response at the upper boundary is shown, but 
 the response is available along the entire enclosing boundary $\setdD$. 
 The time-reversed response $\v(\bx,\bxA,-t)$ is fed to sources (the red dots) at the original positions of the receivers, which emit the wave field back into the medium.
The other panels in column (a) show
``snapshots'' (i.e., wave fields frozen at constant time) of the wave field propagating through the medium.
For negative time ($\cdots$ $-t_2$, $-t_1$ $\cdots$), the field follows the same paths as the original field, but in opposite direction. Then, at $t=0$, the field focuses at the position $\bxA$ of the original source.
Because there is no sink to absorb the focused field, the wave field continues its propagation, away from the focal point. 
Hence, the focal point acts as a  virtual source. The snapshots for positive time ($\cdots$ $+t_1$, $+t_2$ $\cdots$) show the response to this virtual source.
The virtual source is omni-directional and radiates a perfect replica of the original field into the inhomogeneous medium.
Mathematically, time-reversal acoustics is formulated as follows \citep{Derode2003JASA}:
\begin{equation}\label{eq1}
G(\bxB,\bxA,t)+G(\bxB,\bxA,-t)= 2\oint_{\setdD}\underbrace{G(\bxB,\bx,t)}_{\rm``propagator"}*\underbrace{\v(\bx,\bxA,-t)}_{\rm ``secondary\,sources"}{\rm d}\bx
\end{equation}
(see Supplementary Information). 
On the right-hand side, the time-reversed field $\v(\bx,\bxA,-t)$ is propagated through the medium
by the Green's function $G(\bxB,\bx,t)$  from the sources at $\bx$ on the boundary $\setdD$ to any receiver position $\bxB$ inside the medium (the asterisk denotes convolution). 
The integral is taken along all sources $\bx$ on the closed boundary. Note that the right-hand side resembles Huygens' principle, 
which states that each point of an incident wave field acts as a secondary source,  except that here the 
secondary sources on $\setdD$ consist of time-reversed measurements rather than an actual incident field. 
On the left-hand side, the time-reversed Green's function $G(\bxB,\bxA,-t)$ represents the wave field at negative time that converges to  the focal point 
 $\bxA$; the Green's function $G(\bxB,\bxA,t)$ is the response at positive time to the virtual source at $\bxA$.

 Figure \ref{Fig1}(b) shows what happens when the time-reversed response is emitted into the medium by sources (red dots) at the upper boundary only. 
 The field still focuses at $t=0$, but in addition several ghost foci occur at $t=0$. The field at positive time is a virtual-source response, contaminated 
by artefacts, caused by the ghost foci. Moreover, because the focal point is illuminated mainly from above, the virtual source is far from isotropic: it radiates mainly downward.
 
  We now introduce the single-sided focusing approach, which is designed to overcome the limitations of the time-reversal approach in inhomogeneous media with strong impedance contrasts.
   The upper panel in Figure \ref{Fig1}(c) shows a 2D focusing function $F(\bx,\bxA,t)$,
  for the same focal point $\bxA$ as in the time-reversal example. Note that the main event (indicated in blue) is the same as that in $\v(\bx,\bxA,-t)$ in the upper panel in Figure \ref{Fig1}(b),
  but the other events in $F(\bx,\bxA,t)$ come after the main event (instead of preceding it,
 like in $\v(\bx,\bxA,-t)$).
  The snapshots in Figure \ref{Fig1}(c) show the propagation of this focusing function through the medium.
Mathematically, the emission of the focusing function $F(\bx,\bxA,t)$ into the medium by sources at $\bx$ at the upper boundary $\setdD_0$ is described by 
\begin{equation}\label{eq2}
G(\bxB,\bxA,t)+\mbox{anti-symmetric artefacts}=
\int_{\setdDR} G(\bxB,\bx,t)*F(\bx,\bxA,t){\rm d}\bx
\end{equation}
(see Supplementary Information). The right-hand side resembles again Huygens' principle, this time with the focusing function
defining  secondary sources on $\setdDR$ only.
The left-hand side represents the virtual-source response $G(\bxB,\bxA,t)$, contaminated by artefacts that are
anti-symmetric in time. Because the anti-symmetric term vanishes at $t=0$, the panel at $t=0$ in Figure \ref{Fig1}(c) shows a ``clean'' focus. Like in the time-reversal method, 
the focused field acts as a virtual source.
The snapshots at positive time show that this virtual source radiates mainly downward.

Next, we symmetrize both sides of equation (\ref{eq2}), by adding the time-reversal. This suppresses the anti-symmetric artefacts:
\begin{equation}\label{eq3}
G(\bxB,\bxA,t)+G(\bxB,\bxA,-t)=
\mbox{Symmetrize}\Bigl(\int_{\setdDR} G(\bxB,\bx,t)*F(\bx,\bxA,t){\rm d}\bx\Bigr)
\end{equation}
(see Supplementary Information).  Note that the left-hand side is identical to that in equation (\ref{eq1}). However, unlike equation (\ref{eq1}), the right-hand side of equation (\ref{eq3}) 
contains an integral along the accessible boundary $\setdD_0$ only. 
Symmetrization implies 
addition of the snapshots at negative times in Figure \ref{Fig1}(c) to those at the corresponding positive times and vice versa, see  Figure \ref{Fig1}(d). 
Note that these superposed snapshots  are nearly identical to those obtained by emitting the time-reversed response into the medium from the entire enclosing boundary (Figure \ref{Fig1}(a)).
The remaining artefacts are caused by the finite source aperture and the fact that evanescent waves are neglected in equations (\ref{eq2}) and (\ref{eq3}) (see Supplementary Information). 

\section* {Retrieving virtual sources and receivers from single-sided reflection data}

\noindent {\bf Virtual acoustics methodology.}
The snapshots in Figure \ref{Fig1} (for both methods) were obtained by numerically modelling the medium's response to fields emitted from (parts of) its boundary. 
These snapshots nicely visualise the propagation, scattering, focusing and defocusing of the fields inside the medium.
In practical situations these fields are not visible, unless receivers would be placed throughout the medium, which is of course not feasible.
However, our focusing methodology  can be extended to create not only virtual sources, but also virtual receivers  anywhere inside the medium. 
As input we need the reflection response of the medium, 
measured with sources and receivers at the accessible boundary $\setdDR$ only (hence, no physical sources nor receivers are needed inside the medium).
The reflection response is represented by the Green's function $G(\bxr,\bx,t)$, where $\bx$ denotes the variable position of the source and $\bxr$ that of the receiver, both at $\setdDR$.
Consider the following variant of equation (\ref{eq3})
\begin{equation}\label{eq4}
G(\bxB,\bx,t)+G(\bxB,\bx,-t)=
\mbox{Symmetrize}\Bigl(\int_{\setdDR} G(\bxr,\bx,t)*F(\bxr,\bxB,t){\rm d}\bxr\Bigr)
\end{equation}
(see Supplementary Information). This expression shows how the recorded data $G(\bxr,\bx,t)$, measured at the upper boundary of the medium, 
are transformed into $G(\bxB,\bx,t)$ and its time-reversal, being the response to a real source at $\bx$,
observed by  a virtual receiver at $\bxB$ anywhere inside the medium. 
The focusing function $F(\bxr,\bxB,t)$, required for this transformation, can be derived from the recorded data $G(\bxr,\bx,t)$,
using the multidimensional Marchenko method \citep{Wapenaar2013PRL, Slob2014GEO, Wapenaar2016GJI, Brackenhoff2016MSC, Neut2017JASA}.
We have implemented a 2D version of the Marchenko method as an iterative process \citep{Thorbecke2017GEO}. The time-reversal of the direct arriving wave between $\bxr$ and $\bxB$ is used as an initial estimate of the 
focusing function $F(\bxr,\bxB,t)$. This direct arrival, in turn, is based on an estimate of the propagation velocity of the medium.
This does not require information about the layer interfaces, nor about the internal structure of the layers:
 a smooth background model suffices to compute the direct arrival \citep{Broggini2014GEO}. Note that estimating a background model 
is state-of-the-art methodology in geophysical imaging \citep{Harlan2008GEO}. 
Then, by evaluating equation (\ref{eq4}) we obtain $G(\bxB,\bx,t)$ for any virtual receiver position $\bxB$ inside the medium.
Next, using the retrieved virtual-receiver data $G(\bxB,\bx,t)$ in the right-hand side of equation (\ref{eq3}),
we obtain $G(\bxB,\bxA,t)$ and its time-reversal,
being the response to a virtual source at $\bxA$, observed by virtual receivers at $\bxB$.

Theoretical research shows that this methodology can be generalised for vectorial wave fields in lossless media, such as electromagnetic waves, elastodynamic waves
(after decomposition at the surface into $P$- and $S$-waves),
etc. \citep{Wapenaar2014PRE, Costa2014PRE}. Small to moderate propagation losses can be accommodated by applying loss corrections to the data before applying the Marchenko method \citep{Alkhimenkov2017MSC}.

In the following we apply the virtual acoustics methodology for scalar wave fields in lossless media, as outlined above, to ultrasonic physical model data and seismic reflection data. \\

\begin{figure}
\vspace{-14cm}
\centerline{\epsfysize=26 cm \epsfbox{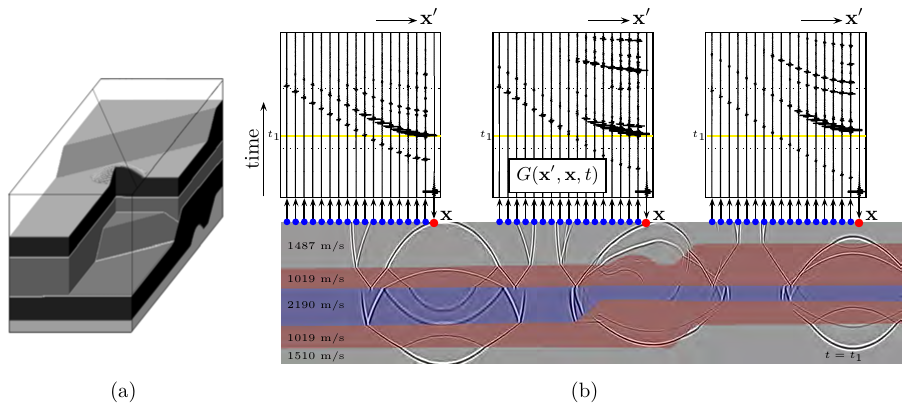}}
\vspace{-5.5cm}
\caption{\small (a) 3D physical model. The grey-levels indicate different propagation velocities and mass densities. Ultrasonic reflection experiments are carried out along the diagonal line above the model.
(b) 2D cross-section of the physical model (with modelled snapshots, for visualisation only) and the actually recorded response at the surface, $G(\bxr,\bx,t)$ 
(here shown for 3 source positions $\bx$ and 3 $\times$ 17 receiver positions $\bxr$).  
}\label{Fig2}
\end{figure}

\noindent {\bf Application to ultrasonic physical model data.}
Figure \ref{Fig2}(a) shows a 3D physical model, composed of silicone gel and beeswax layers with different acoustic propagation velocities 
(their numerical values are tabulated in Figure \ref{Fig2}(b)). The size of the model is $70\times 600 \times 600$ mm. 
The model is placed in a watertank and probed with ultrasound, emitted and received by piezo-electric transducers in the water. 
The acquisition is carried out along a horizontal diagonal line (indicated in Figure \ref{Fig2}(a)),  $12$ mm above the upper boundary of the model and perpendicular to its main structures. 
A 2D cross-section of the model below the acquisition line is shown in Figure \ref{Fig2}(b).
The emitting transducer sends a sweep signal in the frequency range 0.4 MHz to 1.8 MHz. 
The resulting wave field propagates through the water into the model, propagates and scatters inside the model, 
and propagates back through the water to the acquisition line, where it is recorded by a receiving transducer. The recorded response is deconvolved for 
the sweep signal, effectively compressing the source signal to a short zero-phase pulse with a central frequency of 1.1 MHz \citep{Blacquiere97SEG}.
This experiment is repeated 106 times, with the source at the same position
and the receiver moving along the acquisition line in steps of 1.25 mm. Next, the source is moved 1.25 mm along the line and again 106 traces are recorded. This whole process is carried out
301 times, leading to a recorded reflection response consisting of 301 $\times$ 106 =  31 906 traces. Figure \ref{Fig2}(b) shows 51 of those traces, for 3 source positions and 17 receivers per source position. 
Before further processing, source-receiver reciprocity is applied, effectively doubling the number of traces, and
the data are interpolated to a twice as dense spatial grid (source and receiver spacing 0.625 mm) to suppress spatial aliasing.
 \begin{figure}
\vspace{-2cm}
\centerline{\epsfxsize=21 cm \epsfbox{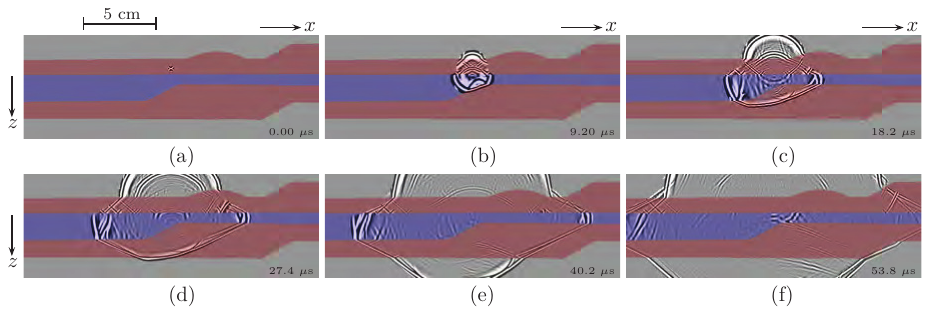}}
\vspace{-20cm}
\caption{\small Virtual response $G(\bxB,\bxA,t)+G(\bxB,\bxA,-t)$, retrieved  from the single-sided ultrasonic reflection response  $G(\bxr,\bx,t)$ of the physical model in Figure \ref{Fig2}(a). 
(a) $t=0$ $\mu$s.
(b) $t=9.2$ $\mu$s.
(c) $t=18.2$ $\mu$s.
(d) $t=27.4$ $\mu$s.
(e) $t=40.2$ $\mu$s.
(f) $t=53.8$ $\mu$s.
}\label{Fig3}
\end{figure}

We denote the recorded reflection response by Green's function $G(\bxr,\bx,t)$, where $\bx$ denotes the variable position of the source and $\bxr$ that of the receiver
(actually the recorded response is the Green's function convolved with the compressed source pulse, but for the sake of simplicity we treat the recorded data as a Green's function). 
We apply the methodology discussed above to this response.
Figure \ref{Fig3} shows snapshots of the virtual acoustic response $G(\bxB,\bxA,t)+G(\bxB,\bxA,-t)$,
for a fixed virtual source inside the second layer of
the 3D physical model and variable virtual receiver positions $\bxB$ throughout the 2D cross-section of the model. The different colours in the background of this figure indicate the different layers. 
We used the velocities of these layers to model the direct arrivals, as initial estimates for the focusing functions.
Note, however, that we did not use information about the layer interfaces for the 
retrieval of the virtual response: all scattering information comes directly from the recorded reflection response.
 The figure clearly shows the evolution of the wave field through the medium, including scattering at the layer interfaces. Imperfections are explained by
 the finite aperture, the limited radiation angles of the piezo-electric transducers,  the negligence of evanescent waves
 and  the fact that we used a 2D method to retrieve this virtual wave field in a 3D medium.\\
\begin{figure}
\vspace{-2cm}
\centerline{\epsfxsize=21 cm \epsfbox{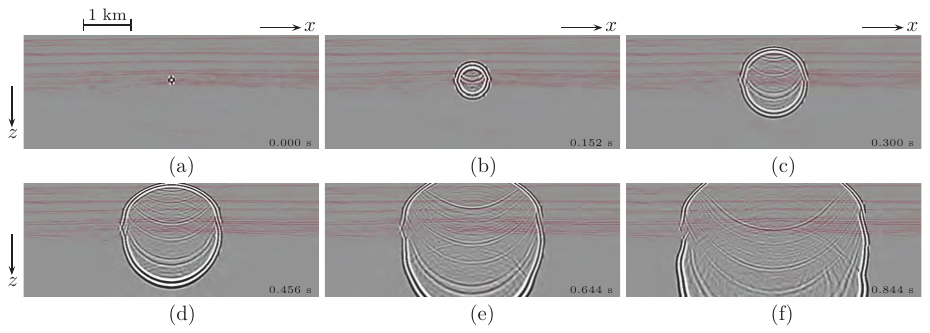}}
\vspace{-20cm}
\caption{\small Virtual response $G(\bxB,\bxA,t)+G(\bxB,\bxA,-t)$, retrieved  from the single-sided seismic reflection response  $G(\bxr,\bx,t)$ of the V\o ring Basin. 
(a) $t=0$ ms.
(b) $t=152$ ms.
(c) $t=300$ ms.
(d) $t=456$ ms.
(e) $t=644$ ms.
(f)  $t=844$ ms.
}\label{Fig4}
\end{figure}

\noindent {\bf Application to seismic reflection data.}
 The proposed methodology can be applied to reflection data at a wide range of scales. 
 Next we apply our methodology to vintage seismic reflection data, 
 acquired in 1994 over the V\o ring Basin by SAGA Petroleum A.S. (currently part of Statoil ASA).
 We use a smooth background model to define the initial estimates of the focusing functions.
 Figure \ref{Fig4} shows snapshots of $G(\bxB,\bxA,t)+G(\bxB,\bxA,-t)$ obtained from these seismic data.
 Again, the evolution of the retrieved wave field clearly includes
  the primary and multiply scattered events, 
 which have been obtained directly from the recorded reflection data. In the background these snapshots show an independently obtained seismic image of the interfaces between the geological layers, for visualisation only.  
 Note the consistency between the position of these interfaces and the apparent origin of scattering in the snapshots.
 
 \section*{Discussion}
 
 The ability to create virtual sources and receivers inside a medium from single-sided reflection data opens new ways for imaging and monitoring.
An exciting new field in medical imaging is photoacoustic (PA) imaging \citep{Wang2016NatureMethods}, a method which employs the conversion of optical energy into acoustic energy at those 
locations inside the medium where light is absorbed.
The resulting acoustic wave field may be very complex because usually many PA sources go off simultaneously and inhomogeneities in the medium may cause reflection artefacts \citep{Singh2016BOE}.
Our proposed virtual acoustics methodology could be applied to ultrasonic reflection measurements to predict the direct and scattered wave fields of (clusters of) virtual PA sources,  
thus improving the interpretation and imaging of the complex wave field of actual PA sources. 
With the emergence of dual-modality ultrasound and photoacoustic imaging tools \citep{Daoudi2014OE} this becomes feasible and the first steps in this direction have already been made \citep{Neut2017JASA}.
Note that in medical applications it is often sufficient to use a homogeneous background model, which means that analytical expressions can be used for the initial estimate of the focusing functions. Real-time application
of our virtual acoustics methodology for medical imaging therefore seems feasible, particularly when the imaging is restricted to a finite region of interest.

 Another exciting potential application is the investigation of 
  induced seismicity. By acquiring high-resolution seismic reflection data in areas prone to induced seismicity, our virtual acoustics approach could 
 forecast the wave field and the associated ground motion caused by possible future earthquakes. Moreover, when the same acquisition system is also used to passively record the response to
 actual induced earthquakes, 
 our method could be used to create virtual seismometers in the subsurface around the actual earthquake and use these to retrieve accurate knowledge of the source mechanism of the earthquake,
  insight in the evolution of the geomechanical state of the subsurface (horizontal and vertical stress distribution, fault and fracture properties etc.), 
  and deep understanding of the link between the earthquake and the observed ground motion. 
 
\section*{Data availability}

The source code that was used to generate Figure \ref{Fig1}, including the Marchenko method, can be downloaded from https://github.com/JanThorbecke/OpenSource.
 The physical model dataset analysed in Figures \ref{Fig2} and \ref{Fig3} are available from the corresponding author on reasonable request.
The seismic reflection data analysed in Figure \ref{Fig4} are available from Statoil ASA, but restrictions apply to the availability of these data, which were used under license for the current study,
 and so are not publicly available. Data are however available from the authors upon reasonable request and with permission of Statoil ASA.

\bibliographystyle{gji}

\mbox{}\\
{\bf Acknowledgements}
The authors thank Statoil ASA for giving permission to use the vintage seismic reflection data of the V\o ring Basin.
This project has received funding from the European Research Council (ERC) under the European Union's Horizon 2020 research 
and innovation programme (grant agreement No: 742703).

\mbox{}\\
{\bf Author contributions}
K.W., E.S. and J.v.d.N. conceived the methodology.
J.B. developed software and applied the methodology to the physical model data (Fig 3) and the vintage seismic data (Fig 4). 
J.T. developed software and conducted the numerical experiment (Fig 1).
E.V. was responsible for preprocessing of the two datasets.
K.W. wrote the paper. 
All authors reviewed the manuscript.




\newpage
\noindent
{\Huge Virtual acoustics in inhomogeneous media with
single-sided access: Supplementary Information}
\appendix

\section{Classical representation of the homogeneous Green's function}

\subsection{Definition of the homogeneous Green's function}

Consider an inhomogeneous lossless acoustic  medium with compressibility $\kappa(\bxb)$ and mass density $\rho(\bxb)$.
Here $\bxb$ denotes the Cartesian coordinate vector $\bxb=(x_1,x_2,x_3)$; the $x_3$ axis is pointing downward.
In this medium a space ($\bxb$) and time ($t$) dependent source distribution $q(\bxb,t)$ is present, with $q$ defined as the volume-injection rate density.
The acoustic wave field, caused by this source distribution, is described in terms of the acoustic pressure
$p(\bxb,t)$ and the particle velocity $v_i(\bxb, t)$.
These field quantities obey the equation of motion and the stress-strain relation, according to
\begin{eqnarray}
\rho\partial_t v_i+\partial_i p&=&0,\label{eqA1}\\
\kappa\partial_t p+\partial_iv_i&=&q.\label{eqA2}
\end{eqnarray}
Here $\partial_t$ and $\partial_i$ stand for the temporal and spatial differential operators $\partial/\partial t$ and $\partial/\partial x_i$, respectively.
The summation convention applies to repeated subscripts.
When $q$ is an impulsive source at  $\bxb=\bxA$ and $t=0$, according to 
\begin{equation}\label{eq2d}
q(\bxb,t)=\delta(\bxb-\bxA)\delta(t),
\end{equation}
 then the causal solution of equations
(\ref{eqA1}) and (\ref{eqA2}) defines the Green's function, hence
\begin{equation}\label{eqA3}
p(\bxb,t)=G(\bxb,\bxA,t).
\end{equation}
By eliminating $v_i$ from equations (\ref{eqA1}) and (\ref{eqA2}) and substituting equations (\ref{eq2d}) and (\ref{eqA3}), we find that 
the Green's function $G(\bxb,\bxA,t)$ obeys the following wave equation
\begin{equation}\label{eqA4}
\partial_i(\rho^{-1}\partial_i G) - \kappa\partial_t^2G=-\delta(\bxb-\bxA)\partial_t\delta(t).
\end{equation}
Wave equation (\ref{eqA4}) is symmetric in time, except for the source on the right-hand side, which is anti-symmetric. Hence, 
the time-reversed Green's function $G(\bxb,\bxA,-t)$ obeys the same wave equation, but with opposite sign for the source.
By summing the wave equations for $G(\bxb,\bxA,t)$ and $G(\bxb,\bxA,-t)$, the sources on the right-hand sides cancel each other, hence, the  function
\begin{equation}\label{eqA8}
G_{\rm h}(\bxb,\bxA,t)=G(\bxb,\bxA,t)+G(\bxb,\bxA,-t)
\end{equation}
obeys the homogeneous equation
\begin{equation}\label{eqA9}
\partial_i(\rho^{-1}\partial_i G_{\rm h}) - \kappa\partial_t^2G_{\rm h}=0.
\end{equation}
Therefore $G_{\rm h}(\bxb,\bxA,t)$, as defined in equation (\ref{eqA8}), is called the homogeneous Green's function.

\subsection{Reciprocity theorems}

We define the temporal Fourier transform of a space- and time-dependent quantity $p(\bxb,t)$ as 
\begin{equation}\label{eqA11}
p(\bxb,\omega)=\int_{-\infty}^\infty p(\bxb,t)\exp(-j\omega t){\rm d}t,
\end{equation}
where $\omega$ is the angular frequency and $j$ the imaginary unit. To keep the notation simple, we denote quantities in the time and frequency domain by the same symbol.
In the frequency domain, equations (\ref{eqA1}) and (\ref{eqA2}) transform to
\begin{eqnarray}
j\omega\rho v_i+\partial_i p&=&0,\label{eqA12}\\
j\omega\kappa p+\partial_iv_i&=&q.\label{eqA13}
\end{eqnarray}
We introduce two independent acoustic states, which will be distinguished by subscripts $A$ and $B$. Rayleigh's reciprocity theorem is obtained by considering the
quantity $\partial_i\{p_Av_{i,B}-v_{i,A}p_B\}$, applying the product rule for differentiation, substituting equations (\ref{eqA12}) and (\ref{eqA13}) for both states,
integrating the result over a spatial domain $\setD$ enclosed by boundary $\setdD$ with outward pointing normal $n_i$, and applying the theorem of Gauss
 \citep{Hoop88JASA, Fokkema93Book}.
Assuming that in $\setD$ the medium parameters $\kappa(\bxb)$ and $\rho(\bxb)$ in the two states are identical, this yields Rayleigh's reciprocity theorem of the convolution type 
\begin{equation}\label{eqA14}
\int_{\setD}\{\pa\qb-\qa\pb\}{\rm d}\bxb=
-\oint_{\setdD}\frac{1}{j\omega\rho}\{\pa\partial_i\pb-(\partial_i\pa)\pb\}n_i{\rm d}\bxb.
\end{equation}
We derive a second form of Rayleigh's reciprocity theorem for time-reversed wave fields.
In the frequency domain, time-reversal is replaced by complex conjugation. When $p$ 
is a solution of equations (\ref{eqA12}) and (\ref{eqA13}) 
with source distribution $q$ (and real-valued medium parameters), then $p^*$
obeys the same equations with source distribution $-q^*$
(the superscript $^*$ denotes complex conjugation).
Making these substitutions for state $A$ in equation (\ref{eqA14}) we obtain Rayleigh's reciprocity theorem of the correlation type  \citep{Bojarski83JASA}
\begin{equation}\label{eqA15}
\int_{\setD}\{\pa^*\qb+\qa^*\pb\}{\rm d}\bxb=
-\oint_{\setdD}\frac{1}{j\omega\rho}\{\pa^*\partial_i\pb-(\partial_i\pa^*)\pb\}n_i{\rm d}\bxb.
\end{equation}

 \subsection{Representation of the homogeneous Green's function}

We choose point sources in both states, according to $q_A(\bxb,\omega)=\delta(\bxb-\bxA)$ and $q_B(\bxb,\omega)=\delta(\bxb-\bxB)$, with $\bxA$ and $\bxB$ both in $\setD$. 
The fields in states $A$ and $B$ are thus expressed in terms of Green's functions, according to
\begin{eqnarray}
p_A(\bxb,\omega)&=&G(\bxb,\bxA,\omega),\label{eqA16}\\
p_B(\bxb,\omega)&=&G(\bxb,\bxB,\omega),\label{eqA16b}
\end{eqnarray}
with $G(\bxb,\bxA,\omega)$ and $G(\bxb,\bxB,\omega)$ being the Fourier transforms of $G(\bxb,\bxA,t)$ and $G(\bxb,\bxB,t)$, respectively.
Making these substitutions in equation (\ref{eqA15}) and using source-receiver reciprocity of the Green's functions gives  \citep{Porter70JOSA, Oristaglio89IP, Wapenaar2004PRL, VanManen2005PRL}
\begin{equation}\label{eqA19}
G_{\rm h}(\bxB,\bxA,\omega)=\oint_{\setdD}\frac{-1}{j\omega\rho(\bxb)}\{\partial_iG(\bxB,\bxb,\omega)G^*(\bxb,\bxA,\omega)
-G(\bxB,\bxb,\omega)\partial_iG^*(\bxb,\bxA,\omega)\}n_i{\rm d}\bxb,
\end{equation}
where $G_{\rm h}(\bxB,\bxA,\omega)$ is the homogeneous Green's function in the frequency domain, defined as
\begin{equation}\label{eqA20}
G_{\rm h}(\bxB,\bxA,\omega)=G(\bxB,\bxA,\omega)+G^*(\bxB,\bxA,\omega)
=2\Re \{G(\bxB,\bxA,\omega)\},
\end{equation}
where $\Re$ denotes the real part. Equation (\ref{eqA19}) is an exact representation for the homogeneous Green's function $G_{\rm h}(\bxB,\bxA,\omega)$.

When $\setdD$ is sufficiently smooth and the medium outside $\setdD$ is homogeneous (with mass density $\rho_0$ and compressibility $\kappa_0$), 
the two terms under the integral in equation (\ref{eqA19}) are nearly identical (but with opposite signs),
hence
\begin{equation}\label{eqA25}
G_{\rm h}(\bxB,\bxA,\omega)=\frac{2}{j\omega\rho_0}\oint_{\setdD}G(\bxB,\bxb,\omega)\partial_iG^*(\bxb,\bxA,\omega)n_i{\rm d}\bxb.
\end{equation}
The main approximation is that evanescent waves are neglected at $\setdD$  \citep{Wapenaar2011GJI}.
Using equations (\ref{eqA12}) and  (\ref{eqA16})  this becomes
\begin{equation}\label{eqA26}
G_{\rm h}(\bxB,\bxA,\omega)= 2\oint_{\setdD}G(\bxB,\bxb,\omega)V^*(\bxb,\bxA,\omega){\rm d}\bxb,
\end{equation}
where $V(\bxb,\bxA,\omega)=v_i(\bxb,\bxA,\omega)n_i$ stands for the normal component of the particle velocity at $\bxb$ on $\setdD$, due to a source at $\bxA$.
In the time domain this becomes
\begin{equation}\label{eqA27}
G(\bxB,\bxA,t)+G(\bxB,\bxA,-t)= 2\oint_{\setdD}G(\bxB,\bxb,t)*V(\bxb,\bxA,-t){\rm d}\bxb,
\end{equation}
where the inline asterisk ($*$) denotes temporal convolution. This is equation (1) in the main paper.

\section{Single-sided Green's function representations}

\begin{figure}
\centerline{\epsfysize=9 cm \epsfbox{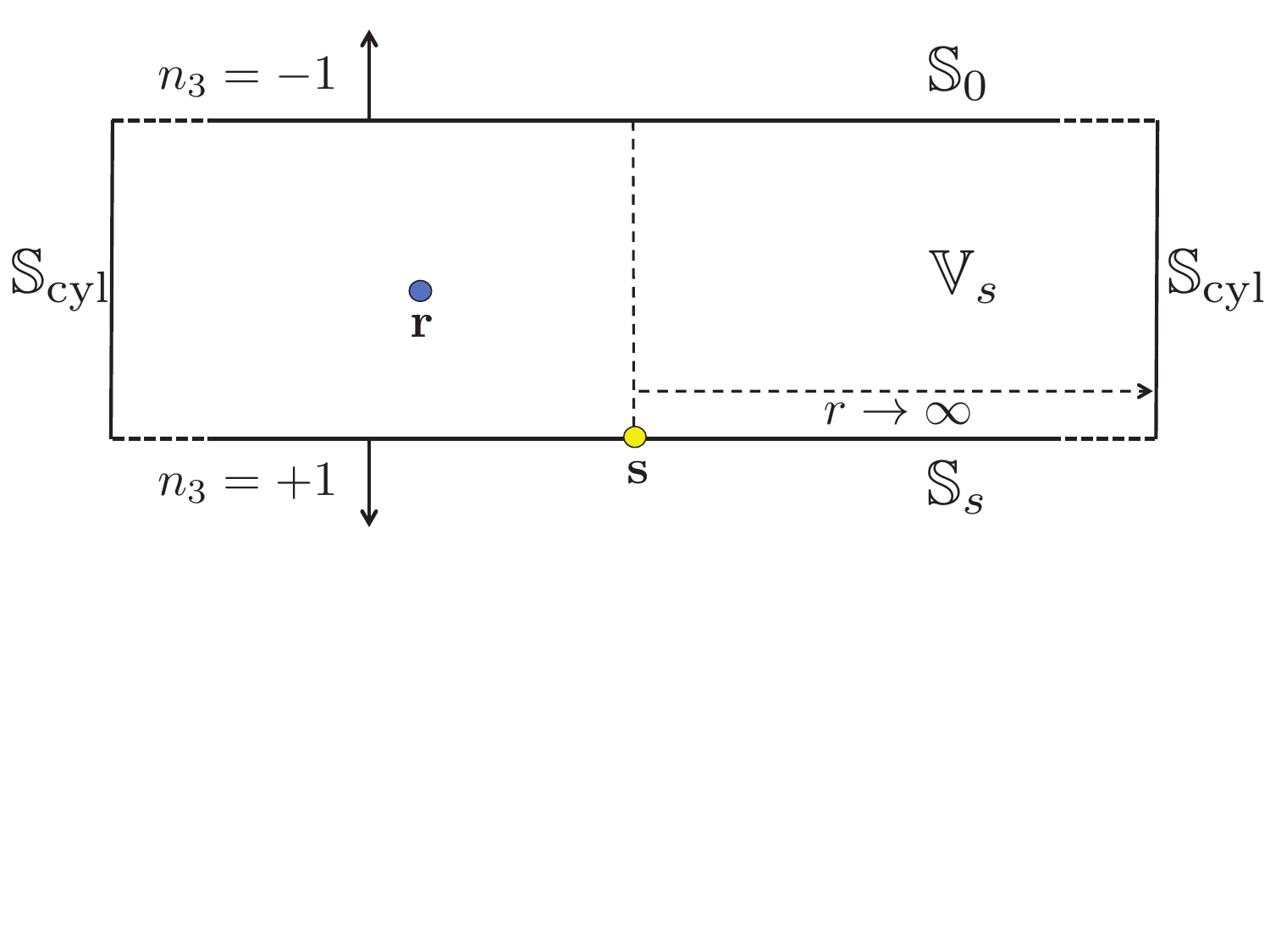}}
\vspace{-3cm}
\noindent{\small {\bf Figure S1.} 
Modified domain $\setD$, enclosed by $\setdDR\cup\setdDA\cup\setdD_{\rm cyl}$ (side view).
}\label{FigS1}
\end{figure}

\subsection{Decomposed reciprocity theorems}

For the derivation of the single-sided Green's function representations  we define  $\setDA$ as the domain enclosed by two horizontal boundaries $\setdDR$ and $\setdDA$, 
and a cylindrical boundary $\setdD_{\rm cyl}$ with infinite radius, see Figure S1.
Here $\setdDR$ is the accessible horizontal boundary of the medium where the measurements take place.
It is defined by $x_3=x_{3,0}$. We assume that the medium above $\setdDR$ is homogeneous.
Furthermore, $\setdDA$ is a horizontal boundary at the depth of $\bxA$ and is defined by $x_3=\xA$.
 The subscript $s$ in $\setdDA$ denotes that this boundary depends on the depth of $\bxA$. Consequently,  $\setDA$ also depends on the depth of $\bxA$. 
 Finally, $\setdD_{\rm cyl}$ is a cylindrical boundary with a vertical axis through $\bxA$ and infinite radius.
 This cylindrical boundary exists between  $\setdDR$ and $\setdDA$  and closes the boundary $\setdD$. 
 
  The contribution of the boundary integral over $\setdD_{\rm cyl}$ in equations  (\ref{eqA14}) and (\ref{eqA15}) vanishes \citep{Wapenaar89Book}.
 This implies that we can restrict the integration to the boundaries  $\setdDR$ and $\setdDA$.
Note that ${\bf n}=(0,0,-1)$ on $\setdDR$ and ${\bf n}=(0,0,+1)$ on $\setdDA$. On the boundaries  $\setdDR$ and $\setdDA$ we decompose the fields in both states into downgoing and upgoing fields, 
according to 
\begin{eqnarray}
\pa&=&\pa^++\pa^-,\\
\pb&=&\pb^++\pb^-,
\end{eqnarray}
 where superscripts $+$ and $-$ stand for downgoing (i.e., propagating in the positive $x_3$-direction) and 
upgoing (i.e., propagating in the negative $x_3$-direction), respectively.  Substituting these expressions into equations (\ref{eqA14}) and (\ref{eqA15}) and using the one-way wave equations
for downgoing and upgoing waves at $\setdDR$ and $\setdDA$, we obtain \citep{Wapenaar89Book}
\begin{equation}\label{eq1S}
\int_{\setDA}\{\pa\qb-\qa\pb\}{\rm d}\bxb=
\int_{\setdDR}\frac{-2}{j\omega\rho}\{(\partial_3\pa^+)\pb^-+(\partial_3\pa^-)\pb^+\}{\rm d}\bxb
+\int_{\setdDA}\frac{2}{j\omega\rho}\{(\partial_3\pa^+)\pb^-+(\partial_3\pa^-)\pb^+\}{\rm d}\bxb,
\end{equation}
and
\begin{equation}\label{eq2S}
\int_{\setDA}\{\pa^*\qb+\qa^*\pb\}{\rm d}\bxb=
\int_{\setdDR}\frac{-2}{j\omega\rho}\{(\partial_3\pa^+)^*\pb^++(\partial_3\pa^-)^*\pb^-\}{\rm d}\bxb
+\int_{\setdDA}\frac{2}{j\omega\rho}\{(\partial_3\pa^+)^*\pb^++(\partial_3\pa^-)^*\pb^-\}{\rm d}\bxb,
\end{equation}
respectively. In the latter equation, evanescent waves at $\setdDR$ and $\setdDA$
are ignored.

\subsection{Single-sided representation of the homogeneous Green's function}

We use  reciprocity theorems (\ref{eq1S}) and (\ref{eq2S}) to derive a single-sided representation of the homogeneous Green's function
 $G_{\rm h}(\bxB,\bxA,\omega)$.
 
For state $A$ we introduce a focusing function $f_1(\bxb,\bxA,\omega)$. Here $\bxA$ denotes the focal point; it lies at $\setdDA$ which we will call the focal plane.
 The focusing function is defined in a source-free truncated medium, 
which is identical to the actual medium above the focal plane $\setdDA$ but homogeneous below this plane, see Figure S2.
For $\bxb$ on the boundaries $\setdDR$ and $\setdDA$ (and in the homogeneous half-spaces above $\setdDR$ and below $\setdDA$), the focusing function is written as a superposition of downgoing and upgoing components, according to
\begin{equation}
f_1(\bxb,\bxA,\omega)=f_1^+(\bxb,\bxA,\omega)+f_1^-(\bxb,\bxA,\omega).
\end{equation}
The downgoing focusing function $f_1^+$ is incident to the inhomogeneous medium from the homogeneous half-space above $\setdDR$, and the upgoing function $f_1^-$ is its response. 
The focusing function propagates and scatters in the
truncated medium in $\setDA$, focuses at $\bxA$ on $\setdDA$, 
and continues as a downgoing wave field $f_1^+$ in the homogeneous lossless half-space below $\setdDA$.
The focusing conditions at the focal plane $\setdDA$ are defined as
\begin{equation}
[\partial_3f_1^+(\bxb,\bxA,\omega)]_{x_3=\xA}=-\half j\omega\rho(\bxA)\delta(\bxh-\bxha),\label{eqS11}
\end{equation}
\begin{equation}
[\partial_3f_1^-(\bxb,\bxA,\omega)]_{x_3=\xA}=0.\label{eqS12}
\end{equation}
Here $\bxh$ and $\bxha$ stand for the horizontal coordinates of $\bxb$ and $\bxA$, respectively.

\begin{figure}
\centerline{\epsfysize=9 cm \epsfbox{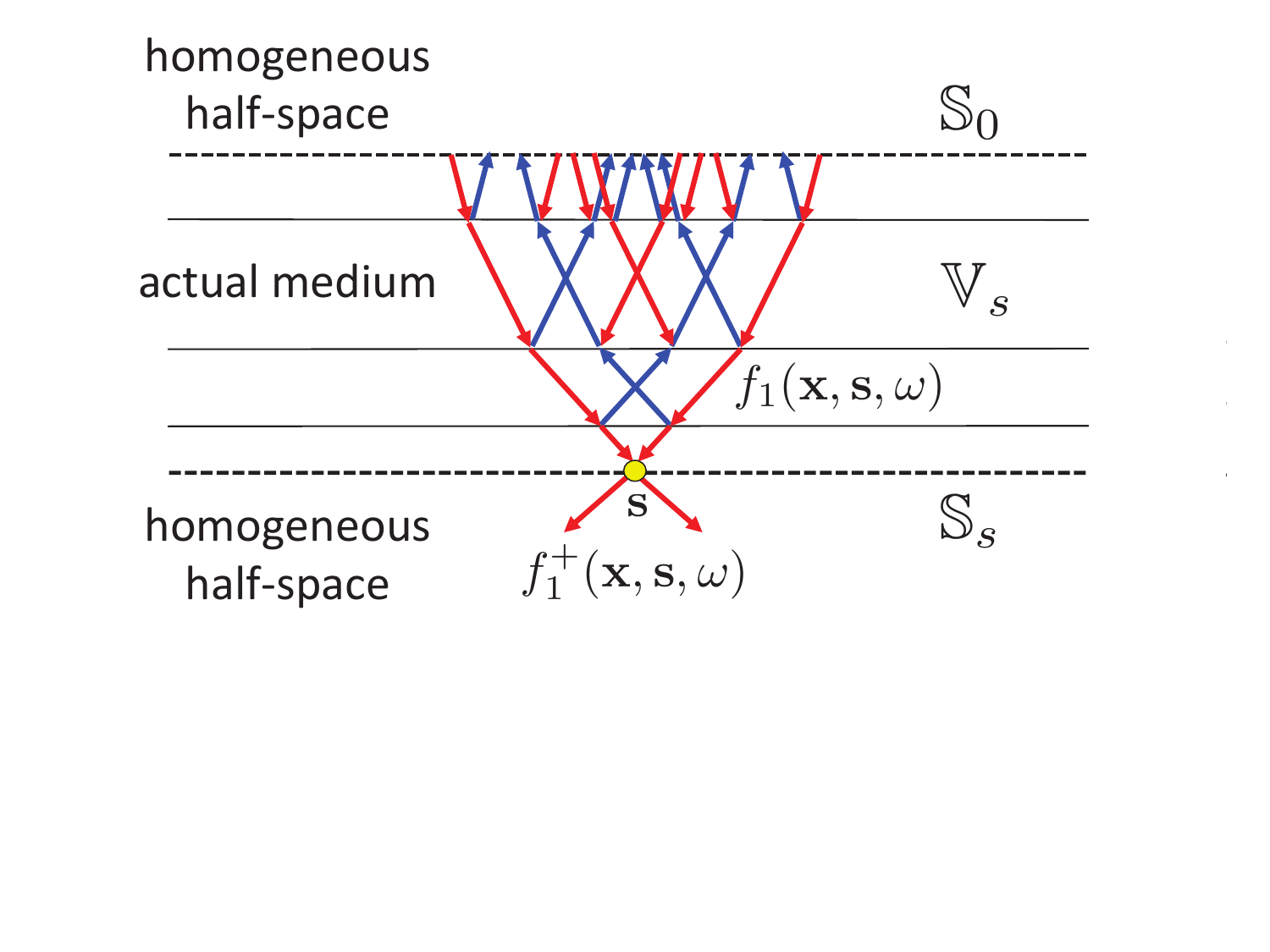}}
\vspace{-2cm}
\noindent{\small {\bf Figure S2.} 
Illustration of the focusing function $f_1(\bxb,\bxA,\omega)$, defined in a truncated version of the actual medium.
}\label{FigS2}
\end{figure}

For state $B$ we take again the Green's function $G(\bxb,\bxB,\omega)$, which, for $\bxb$ at $\setdDR$ and $\setdDA$, is written as 
\begin{equation}
G(\bxb,\bxB,\omega)=G^+(\bxb,\bxB,\omega)+G^-(\bxb,\bxB,\omega).
\end{equation}
Here $\bxB$ can be chosen anywhere below the upper boundary $\setdDR$. 
When $\bxB$ lies above $\bxA$  (and below $\setdDR$, as illustrated in Figure S1), it is by definition situated in $\setDA$. When $\bxB$ lies below $\bxA$, it is 
situated outside $\setDA$. Because the upper half-space above $\setdDR$ is homogeneous, we have
\begin{equation}\label{eqS13}
[G^+(\bxb,\bxB,\omega)]_{x_3=x_{3,0}}=0.
\end{equation}
 Substituting $\pa(\bxb,\omega)=f_1(\bxb,\bxA,\omega)$, $\pa^\pm(\bxb,\omega)=f_1^\pm(\bxb,\bxA,\omega)$, $\qa(\bxb,\omega)=0$, 
 $\pb^\pm(\bxb,\omega)=G^\pm(\bxb,\bxB,\omega)$ and $\qb(\bxb,\omega)=\delta(\bxb-\bxB)$ into equations (\ref{eq1S}) and (\ref{eq2S}), 
using equations (\ref{eqS11}),  (\ref{eqS12}) and  (\ref{eqS13}), gives
\begin{equation}\label{eqS101}
G^-(\bxA,\bxB,\omega)+\chia(\bxB)f_1(\bxB,\bxA,\omega)=
-\int_{\setdDR}\frac{2}{j\omega\rho(\bxb)}\{\partial_3f_1^+(\bxb,\bxA,\omega)\}G^-(\bxb,\bxB,\omega){\rm d}\bxb
\end{equation}
and
\begin{equation}\label{eqS102}
G^+(\bxA,\bxB,\omega)-\chia(\bxB)f_1^*(\bxB,\bxA,\omega)=
\int_{\setdDR}\frac{2}{j\omega\rho(\bxb)}\{\partial_3f_1^-(\bxb,\bxA,\omega)\}^*G^-(\bxb,\bxB,\omega){\rm d}\bxb,
\end{equation}
where $\chia$ is the characteristic function of the domain $\setDA$. It is defined as
\begin{equation}\label{eqC3.2}
\chia(\bxB)=
\begin{cases}
1,   &\text{for } \bxB\text{ in }{\setDA}, \\
\half, &\text{for } \bxB\text{ on }{\setdD=\setdDR\cup\setdDA},\\
 0,  &\text{for } \bxB\text{ outside }{\setDA\cup\setdD}.
 \end{cases}
\end{equation}
Summing  equations (\ref{eqS101}) and (\ref{eqS102}), using $G(\bxb,\bxB,\omega)=G^-(\bxb,\bxB,\omega)$ for $\bxb$ at $\setdDR$, and using source-receiver reciprocity
for the Green's functions, yields
\begin{equation}\label{eqS201}
G(\bxB,\bxA,\omega)+\chia(\bxB)2j\Im\{f_1(\bxB,\bxA,\omega)\}
=\int_{\setdDR} G(\bxB,\bxb,\omega)F(\bxb,\bxA,\omega){\rm d}\bxb,
\end{equation}
with
\begin{equation}
F(\bxb,\bxA,\omega)=-\frac{2}{j\omega\rho(\bxb)}\partial_3\bigl(f_1^+(\bxb,\bxA,\omega)-\{f_1^-(\bxb,\bxA,\omega)\}^*\bigr),
\end{equation}
where $\Im$ denotes the imaginary part.  
Inverse Fourier transforming equation (\ref{eqS201})  gives
\begin{equation}\label{eqS201t}
G(\bxB,\bxA,t)+\chia(\bxB)\{f_1(\bxB,\bxA,t)-f_1(\bxB,\bxA,-t)\}
=\int_{\setdDR} G(\bxB,\bxb,t)*F(\bxb,\bxA,t){\rm d}\bxb.
\end{equation}
This is equation (2) in the main paper.
Taking two times the real part of both sides of equation (\ref{eqS201})  gives
\begin{equation}\label{sup10}
G_{\rm h}(\bxB,\bxA,\omega)=
2\Re\int_{\setdDR} G(\bxB,\bxb,\omega)F(\bxb,\bxA,\omega){\rm d}\bxb.
\end{equation}
In the time domain this becomes
\begin{equation}\label{sup10b}
G(\bxB,\bxA,t)+G(\bxB,\bxA,-t)=
\int_{\setdDR} G(\bxB,\bxb,t)*F(\bxb,\bxA,t){\rm d}\bxb + \int_{\setdDR} G(\bxB,\bxb,-t)*F(\bxb,\bxA,-t){\rm d}\bxb.
\end{equation}
This is equation (3) in the main paper.

Note that the Green's function $G(\bxB,\bxb,\omega)$ on the right-hand side of equation (\ref{sup10}) can be obtained from a similar representation. 
To see this, replace in the right-hand side of equation  
(\ref{sup10}) $\setdDR$ by $\setdDR'$ just above $\setdDR$, replace $\bxb$ on $\setdDR$ by $\bxb'$ on $\setdDR'$, $\bxB$ inside the medium by $\bxb$ on $\setdDR$ and $\bxA$ by $\bxB$.
This gives a representation for $G_{\rm h}(\bxb,\bxB,\omega)$. Using source-receiver reciprocity
we finally get
\begin{equation}\label{sup11}
G_{\rm h}(\bxB,\bxb,\omega)=
2\Re\int_{\setdDR'} G(\bxb',\bxb,\omega)F(\bxb',\bxB,\omega){\rm d}\bxb'.
\end{equation}
In the time domain this becomes
\begin{equation}\label{sup11b}
G(\bxB,\bxb,t)+G(\bxB,\bxb,-t)=
\int_{\setdDR'} G(\bxb',\bxb,t)*F(\bxb',\bxB,t){\rm d}\bxb' + \int_{\setdDR'} G(\bxb',\bxb,-t)*F(\bxb',\bxB,-t){\rm d}\bxb'.
\end{equation}
This is equation (4) in the main paper, where for simplicity the prime in $\setdDR'$ is dropped.

\end{document}